\documentclass[%
 aip,
 amsmath,amssymb,
]{revtex4-1}

\usepackage{graphicx}% Include figure files
\usepackage{dcolumn}% Align table columns on decimal point
\usepackage{bm}% bold math
\usepackage[utf8]{inputenc}
\usepackage[T1]{fontenc}
\usepackage{mathptmx}
\usepackage{etoolbox}
\usepackage{color}

\makeatletter
\def\@email#1#2{%
 \endgroup
 \patchcmd{\titleblock@produce}
  {\frontmatter@RRAPformat}
  {\frontmatter@RRAPformat{\produce@RRAP{*#1\href{mailto:#2}{#2}}}\frontmatter@RRAPformat}
  {}{}
}%
\makeatother
\begin{document}

\preprint{AIP/123-QED}

\title[2D PIC simulation of collective Thomson scattering in a beam-plasma system]{Two-dimensional PIC simulation of collective Thomson scattering in a beam-plasma system}
% Force line breaks with \\

\author{Yuma Sato}
\affiliation{Interdisciplinary Graduate School of Engineering Sciences, Kyushu University \\
6-1 Kasuga-Koen, Kasuga, Fukuoka 816-8580, Japan}

\author{Shuichi Matsukiyo}
\affiliation{Faculty of Engineering Sciences, Kyushu University \\
6-1 Kasuga-Koen, Kasuga, Fukuoka 816-8580, Japan}
\affiliation{Quantum and Spacetime Research Institute (QuaSR), Kyushu University \\
744 Motooka, Nishi-ku, Fukuoka 819-0395, Japan}
\affiliation{International Research Center for Space and Planetary Environmental Science (i-SPES), Kyushu University \\
744 Motooka, Nishi-ku, Fukuoka 819-0395, Japan}
\affiliation{Institute of Laser Engineering, Osaka University \\
2-6, Yamadaoka, Suita, Osaka 565-0871, Japan}
\email{matsukiy@esst.kyushu-u.ac.jp}

%\author{A. Author}
%\author{B. Author}%
% \email{Second.Author@institution.edu.}
%\affiliation{
%Authors' institution and/or address%\\This line break forced with \textbackslash\textbackslash
% \homepage{http://www.Second.institution.edu/~Charlie.Author.}
%\affiliation{%
%Second institution and/or address%\\This line break forced% with \\
%}%
%
\date{\today}% It is always \today, today,

\begin{abstract}
Collective Thomson scattering (CTS) in a beam-plasma system is reproduced by using 2D PIC simulations and the characteristics of the scattered wave spectrum are examined. By formulating the geometric shape of the scattered wave spectrum in wave number space, where the velocity vector of the beam component and the wave vectors of the incident and scattered waves are arbitrary, it is demonstrated that the spectrum in 2D wave number space becomes asymmetric. The spectrum of scattered waves propagating in a specific direction is presented as a function of wavelength to show that the electron (ion) feature is amplified and becomes asymmetric or distorted when Buneman (ion acoustic) instability occurs. An additional simulation is conducted for a weak, linearly stable beam–plasma system with a hot beam, and confirmed that the obtained scattered wave spectrum shows asymmetric feature. The results are expected to be applicable to the interpretation of radar observations of ionospheric plasmas as well as CTS measurements in laboratory plasmas.
\end{abstract}

\maketitle

%%%%%%%%%%%%%%%%%%%%%%%%%%%%%%%%%%%%%%%%
\section{\label{sec:intro}Introduction}
%%%%%%%%%%%%%%%%%%%%%%%%%%%%%%%%%%%%%%%%

Thomson scattering is used for plasma diagnostics. In laboratory
experiments, a laser is injected into the plasma, and the scattered
waves are measured to estimate the plasma parameters based on the
characteristics of the spectrum.
For diagnosing ionospheric plasma,
radio waves are used as the incident waves.
Thomson scattering in a plasma can be classified into two types
depending on the wavelength of the electromagnetic waves.
When the wavelength is shorter than the Debye length, it is
called noncollective scattering, whereas it is called collective
scattering when it is longer. The former results from
scattering by thermally fluctuating free electrons, while the
latter is caused by electrons reflecting the collective motion
of the plasma \cite{froula11}.

The collective Thomson scattering (CTS) in non-equilibrium plasmas
is not yet fully understood, and a comprehensive scattering theory
has not been established. Observations of ionospheric plasma using
the incoherent scatter radar (ISR) have captured scattered wave spectra that cannot
be explained by equilibrium plasma theory \cite{rietveld91,wahlund92a,wahlund92b,forme01,stromme05,ogawa11,schlatter14,akbari17}.
Moreover, in recent power laser experiments on collisionless
shocks \cite{vvv25,matsukiyo22,yamazaki22,schaeffer19,rinderknecht18,lebedev14}
as well as counter-streaming plasma
 \cite{sakai20,morita19,sakawa17,morita13,ross13,ross12,park12},
complex scattered wave spectra have also been reported.

To extract meaningful information from such complex CTS spectra,
attempts have been made to reproduce CTS spectrum using particle-in-cell (PIC)
simulations. Diaz et al. (2008) \cite{diaz08} firstly reproduced CTS
in an equilibrium plasma. It is well known that the scattered wave
spectrum is closely related to the spectral density function,
$
S({\bf k}_S, \omega)=\lim_{V,T \rightarrow \infty}
\left< |N_e(k_S, \omega)|^2 / n_{e0} \right>/VT,
$
which is the ensemble average of the squared electron density
fluctuation spectrum, $N_e(k_S, \omega)$. Here, $V$ is the volume of the plasma, $T$ the observation time, $n_{e0}$ the average electron density, and ${\bf k}_S$ is the scattered wave vector, respectively. They reproduced
$S({\bf k}_S, \omega)$ by performing simulations of
a thermal plasma and calculating the Fourier spectrum of the
electron density fluctuations. Through long time 2D simulations,
by averaging over various wave propagation angles for a specific
$|{\bf k}_S|$ to increase the number of ensembles sufficiently,
they showed that the simulation results were consistent with
the theoretical predictions. Subsequently, they performed electrostatic
2D PIC simulations for electron beam-plasma
systems, showing that asymmetric or distorted intense scattered wave spectra
could be obtained \cite{diaz11,diaz12}.
\textcolor{black}{
Here, the terms "asymmetric" or "distorted" mean that the scattered wave spectrum exhibits a qualitatively asymmetric shape compared to that in an equilibrium plasma.
}
Their approach is
characterized by calculating $S({\bf k}_S, \omega)$
directly from the simulation without simulating the
scattering process.

Sakai et al.(2020,2023) \cite{sakai20,sakai23} recently proposed a
different approach. They
reproduced the two-stream instability using 1D
PIC simulations and the obtained information
of electron and ion density fluctuations (and the incident wave)
is used as input to solve the electromagnetic wave equation to
simulate CTS. They revealed that resultant scattered wave spectra
are highly intense and asymmetric or distorted.
It should be noted that, because it is a one-dimensional
simulation, the scattering angles in terms of the incident
wave are limited to $0^{\circ}$ and $180^{\circ}$. Also,
in this approach, the absorption of the incident wave by the
plasma and the effect of finite size of incident wave packet are
not considered.

In this study, we self-consistently reproduce the long time evolution
of the beam-plasma system and CTS using 2D PIC simulations.
\textcolor{black}{Here, self-consistent means that the evolution of waves in the plasma (both those arising from beam instabilities and those involved in CTS) and the motion of particles (electrons and ions) are solved simultaneously while accounting for their mutual interactions.}
With
actual experimental and observational measurement systems in mind,
we discuss the spectrum of scattered waves with a finite scattering
angle relative to the incident wave (and a beam velocity). In
particular, we investigate how changes in the characteristics of
instability due to differences in beam parameters are reflected
in the multidimensional scattered wave spectrum. %Furthermore,

The paper is organized as follows. Simulation settings are explained in section \ref{sec:setting}. Section \ref{sec:s-structure} shows the simulation results including 2D CTS spectra and the spectra of the scattered waves propagating in a specific direction for various beam parameters. Then in section \ref{sec:weakbeam}, we reproduce the CTS in a weak beam-plasma system. Finally, discussions and summary are given in section \ref{sec:discussion}.

%%%%%%%%
\section{\label{sec:setting}Simulation settings}
%%%%%%%%

%>>>>>>>>>>>>>>>>>>>>>>>>>>>>>>>>>>>>>>>>>>>>>>>>>>
\begin{figure}[ht]
\includegraphics[clip, width=0.95\columnwidth]{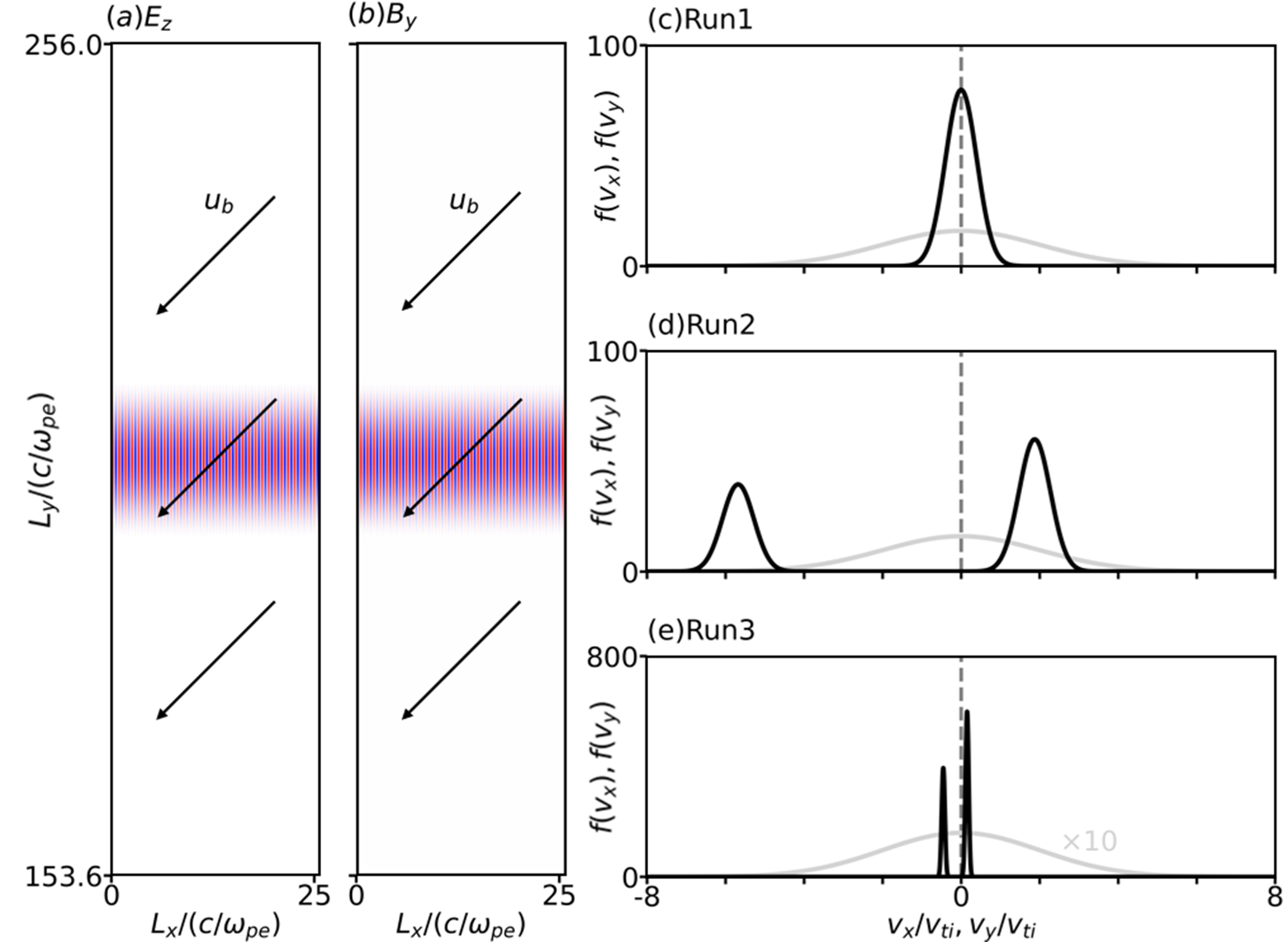}% Here is how to import EPS art
\caption{\label{fig:setting}Initial conditions of simulations. (a) $E_z$
and (b) $B_y$ components of incident wave packet (color scale) and beam
direction (arrows).
Ion (black lines) and electron (gray lines) distribution functions of
(c) Run 1\textcolor{black}{: two-component equilibrium plasma},
(d) Run 2\textcolor{black}{: strong ion-beam plasma}, and
(e) Run 3\textcolor{black}{: weak ion-beam plasma}. \textcolor{black}{Note that the electron distribution
in panel (e) is scaled by a factor of 10 due to the large difference in
peak values between ions and electrons.}}
\end{figure}
%>>>>>>>>>>>>>>>>>>>>>>>>>>>>>>>>>>>>>>>>>>>>>>>>>>
%

Here, we use a 2D periodic boundary \textcolor{black}{electromagnetic} full-PIC simulation of
a system consisting of background ions, electrons, and beam
ions. We further inject a packet of electromagnetic wave into
the system to reproduce the CTS in the beam-plasma system.

The number of spatial grids is 1,024 $\times$ 16,384, with 32
superparticles per species per cell. The grid spacing
corresponds to the Debye length, which translates to a system
size of $L_x/(c/\omega_{pe}) \times L_y/(c/\omega_{pe}) =$
25.6 $\times$ 409.6, where $c$ and $\omega_{pe}$ denote the
speed of light and electron plasma frequency. The time step
is $\omega_{pe} \Delta t = 1.75 \times 10^{-2}$. The
ion-to-electron mass ratio is $m_i/m_e=25$.

As in Fig.\ref{fig:setting}, the incident electromagnetic wave
is a linearly polarized wave packet with monochromatic frequency
propagating in the $x-$direction along $y=L_y/2$.
The wave packet size is set to $\Delta/(c/\omega_{pe})=20$.
Its maximum intensity is $E^2_I/8 \pi n_{e} T_e = 2.0$, wavenumber
is $k_I c/ \omega_{pe}=7.85$, and frequency is
$\omega_I/\omega_{pe}=7.91$, where $n_{e}$ and $T_e$ are electron
density and temperature.

The simulation is performed in the electron rest frame. The
electrons follow a Maxwellian distribution with thermal
velocity of $v_{te}=(T_e/m_e)^{1/2}$,
while the background ions and beam ions follow a
shifted-Maxwellian distribution. The direction of beam drift
velocity is at an angle of $-135^{\circ}$ relative to the $x-$axis.
The background ions are also given a drift velocity to cancel the
net current.

In the next section, we first present the results of three runs.
In Run 1, a two-component equilibrium plasma without a beam is
considered, where the ion temperature is equal to the electron
temperature ($T_i=T_e$). In Run 2, a beam having a relative
density of $n_b/n_e=1/4$ and a temperature of $T_b/T_e=1/4$ is
added. The beam drift velocity is set to $u_b/v_{te}=4$. In
Run 3, the beam drift velocity is set to $u_b/v_{te}=0.32$,
and both the beam and background ion temperatures are reduced
to $T_b=T_i=T_e/100$. The corresponding distribution functions
are depicted in Fig.\ref{fig:setting}(c-e).

%----------
\section{\label{sec:s-structure}2D spectrum of CTS}
%----------
%
%----------
\subsection{Overview}
%----------
%>>>>>>>>>>>>>>>>>>>>>>>>>>>>>>>>>>>>>>>>>>>>>>>>>>
\begin{figure}[ht]
\includegraphics[clip, width=1.0\columnwidth]{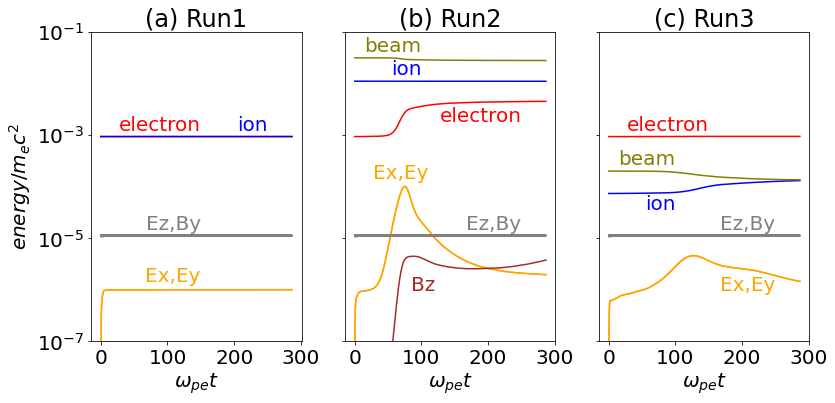}% {fig04.jpg}%Here is how to import EPS art
\caption{\label{fig:Ehistory}Energy time history of (a) Run 1,
(b) Run 2, and (c) Run 3.}
\end{figure}
%>>>>>>>>>>>>>>>>>>>>>>>>>>>>>>>>>>>>>>>>>>>>>>>>>>
%
Fig.\ref{fig:Ehistory} shows the time evolution of the energy
densities of various physical quantities for each run. $E_z$
and $B_y$ are the electromagnetic field components of the
incident wave.

%
%>>>>>>>>>>>>>>>>>>>>>>>>>>>>>>>>>>>>>>>>>>>>>>>>>>
\begin{figure}[ht]
\includegraphics[clip, width=1.0\columnwidth]{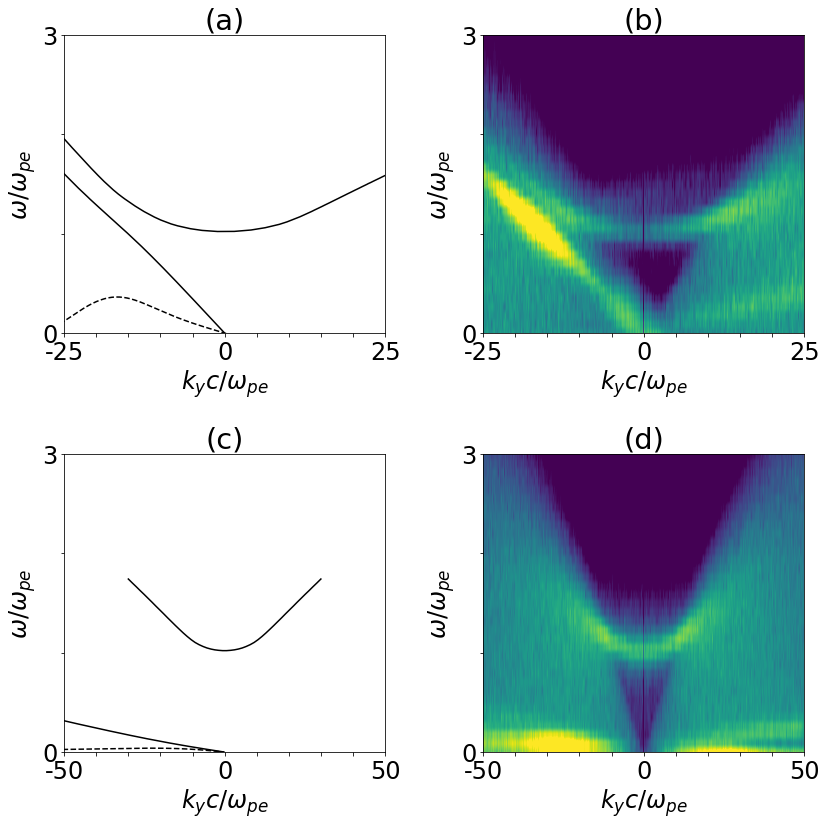}% {fig04.jpg}%Here is how to import EPS art
\caption{\label{fig:disp}Linear dispersion relation of (a) Run 2 and
(c) Run 3. The solid and dashed lines denote real and imaginary parts
of frequency. $\omega-k_y (k_x=0)$ spectra of $E_y$ in the time
range $0 < \omega_{pe}t < 71$ for (b) Run 2 and (d) Run 3.}
\end{figure}
%>>>>>>>>>>>>>>>>>>>>>>>>>>>>>>>>>>>>>>>>>>>>>>>>>>
%

In Run 1, there is little change in energy, indicating that
plasma heating by the incident wave is negligible. According
to linear analysis, Run 2 and Run 3 are expected to exhibit
Buneman instability and ion acoustic instability, respectively.
Fig.\ref{fig:disp}(a) and \ref{fig:disp}(c) depict the linear
dispersion relation for Run 2 and Run 3\textcolor{black}{, which
are obtained by numerically solving the following formula\cite{gary05}.
\begin{equation}
1-\sum_j {Z'(\zeta_j) \over 2 k^2 \lambda_{Dj}}=0
\end{equation}
Here, $Z'$ is the first order derivative of plasma dispersion
function, $\zeta_j=(\omega-ku_j)/\sqrt{2}kv_{tj}$, and
$\lambda_{Dj}$ denotes the Debye length of the $j-$th species.
}
The solid lines
represent \textcolor{black}{the real part of} wave frequency, while the dashed lines denote growth rate \textcolor{black}{(its imaginary part)}.
Fig.\ref{fig:disp}(b) and \ref{fig:disp}(d) show $\omega-k_y$
spectrum obtained by performing a spatio-temporal Fourier transform
on the $E_y$ component for each run in the time range
$0 < \omega_{pe}t < 71$, extracted along $k_x=0$.
The amplification of the $E_x$ and $E_y$ components observed
in Fig.\ref{fig:Ehistory}(b) and \ref{fig:Ehistory}(c) is
attributed to these instabilities.
In Run 2 (Fig.\ref{fig:Ehistory}(b)), although weak, the $B_z$
component is amplified due
to the Weibel instability, which is not the focus of this study.

\begin{figure*}[ht]
\includegraphics[clip, width=\textwidth]{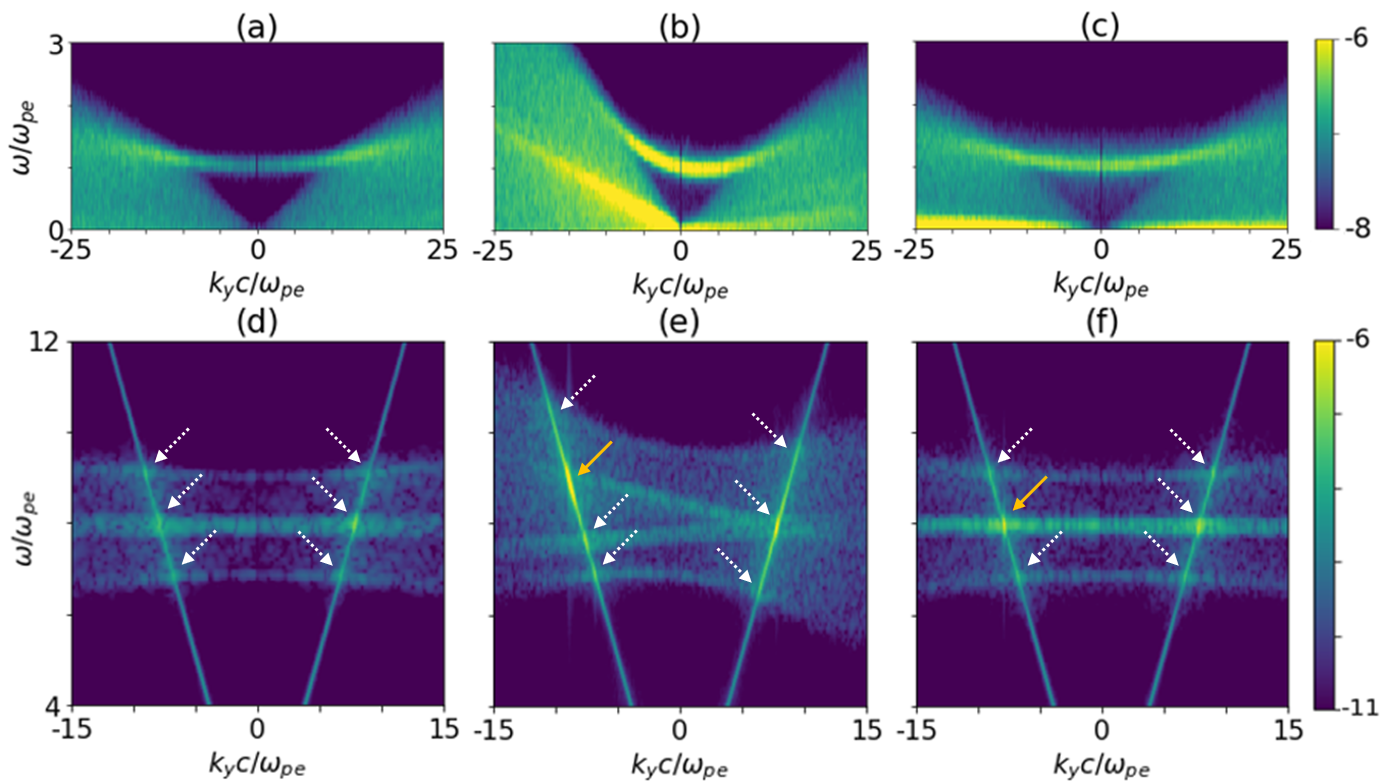}
\caption{\label{fig:wky}$\omega-k_y (k_x=0)$ spectra in the time range
$216 < \omega_{pe}t < 287$ for (a,d) Run 1, (b,e) Run 2, and (c,f) Run3.
The upper and lower panels show $E_y$ and $E_z$ components.}
\end{figure*}

Fig.\ref{fig:wky} shows the similar $\omega-k_y$ spectrum along
$k_x=0$ on the $E_y$ and
$E_z$ components for each run in the later time range
$216<\omega_{pe}t<287$. In Run 1
(Fig.\ref{fig:wky}(a)), Langmuir waves are clearly observed in
$E_y$ spectrum. The weak spectrum in the low-frequency
($\omega \sim 0$) region
corresponds to ion acoustic waves. The $E_z$ component reflects
these waves, showing scattered waves. There are three distinct
bright regions \textcolor{black}{of frequency} along the dispersion relation of light in an
equilibrium plasma \textcolor{black}{indicated by the white dotted arrows}. One is near the frequency of the incident
wave ($\omega_I/\omega_{pe}=7.91$), while the other two are shifted by approximately the plasma frequency to the higher and lower
frequency sides. The former is due to ion acoustic wave scattering,
and the latter are due to Langmuir wave scattering. In the
theory of CTS, the former is called
the ion feature, while the latter is called the electron feature.
In Run 2 (Fig.\ref{fig:wky}(b)), as seen in $E_y$, the beam mode
is strongly amplified due to Buneman instability, and accordingly,
strong scattering due to this beam mode is also observed in $E_z$.
\textcolor{black}{The strongest signal in the scattered wave spectrum
is indicated by the orange arrow (Other scattered signals are indicated by the white dotted arrows)}.
In Run 3 (Fig.\ref{fig:wky}(c)), ion
acoustic instability amplifies low-frequency ion acoustic waves
in the $E_y$ component, and in response, scattered waves at the
frequency near the incident wave are prominently observed in the
$E_z$ component. \textcolor{black}{Again, the strongest scattered wave signal is indicated by the orange arrow and other scattered signals are by the white dotted arrows.}

%
%>>>>>>>>>>>>>>>>>>>>>>>>>>>>>>>>>>>>>>>>>>>>>>>>>>
%>>>>>>>>>>>>>>>>>>>>>>>>>>>>>>>>>>>>>>>>>>>>>>>>>>
%

%----------
\subsection{2D CTS spectrum}
%----------

%
%>>>>>>>>>>>>>>>>>>>>>>>>>>>>>>>>>>>>>>>>>>>>>>>>>>
\begin{figure}[ht]
\includegraphics[clip, width=1.0\columnwidth]{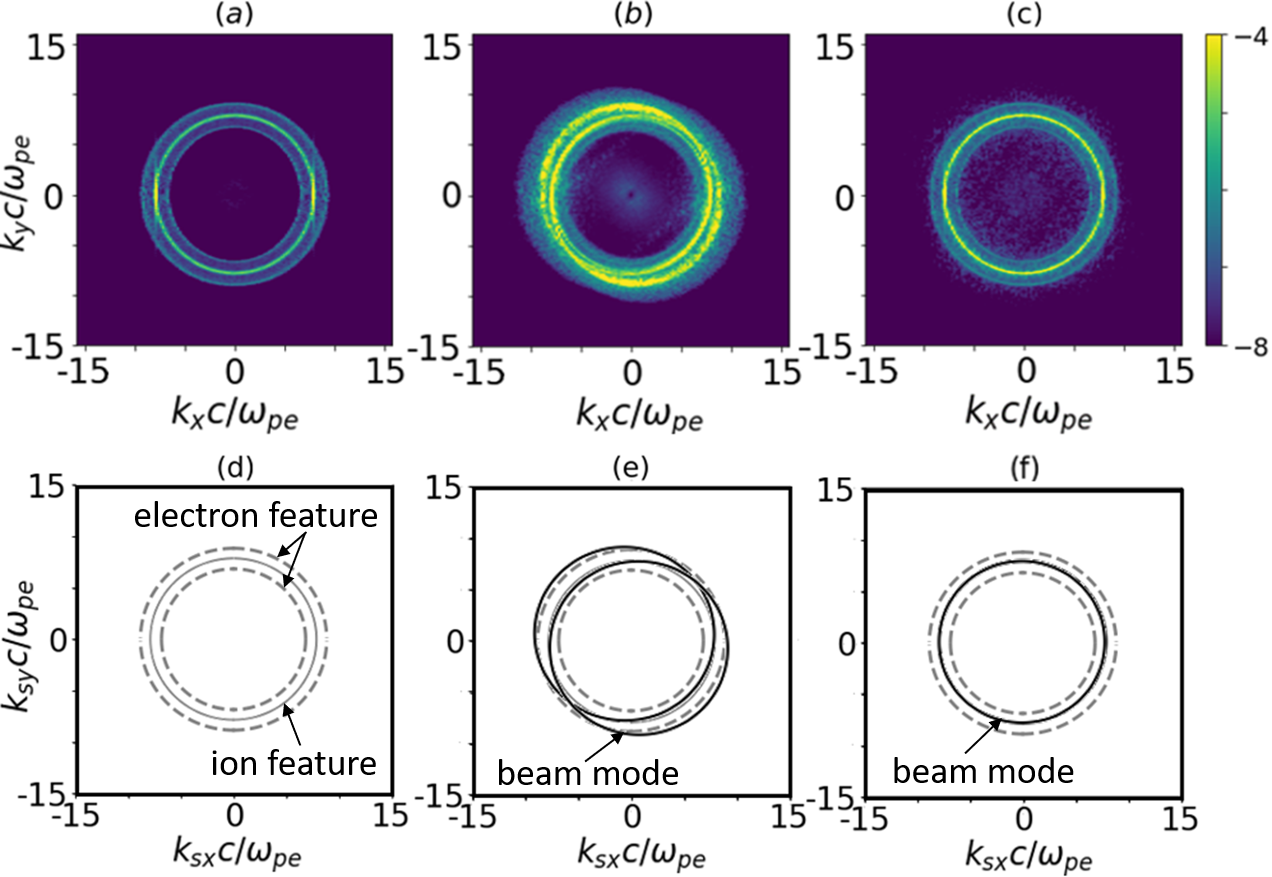}% {fig04.jpg}%Here is how to import EPS art
\caption{\label{fig:kxky}Time averaged $k_x-k_y$ spectra of $E_z$ for
(a) Run 1, (b) Run 2, (c) Run 3, and (d,e,f) their interpretation.}
\end{figure}
%>>>>>>>>>>>>>>>>>>>>>>>>>>>>>>>>>>>>>>>>>>>>>>>>>>
%

By performing a Fourier transform of the $E_z$ component data
at each time step in the $x-$ and $y-$directions, a
$k_x-k_y$ spectrum is obtained. Fig.\ref{fig:kxky} shows
the time-averaged $k_x-k_y$ spectrum obtained over the time
range $200 < \omega_{pe}t < 270$ of the simulation for the
three runs. Any cross-section of these figures represents the
spectrum of scattered light in that direction. Note that
due to the symmetry of the Fourier transform, the spectrum is symmetric about the origin.

As seen in Fig.\ref{fig:kxky}, the characteristic feature of
the $k_x-k_y$ spectrum is the presence of multiple ring structures.
To understand these structures, we consider the resonance
conditions among the incident wave $(\omega_I, {\bm k_I})$,
the scattered waves $(\omega_S, {\bm k_S})$, and the waves in
the plasma $(\omega, {\bm k})$. The frequency and wavenumber
resonance conditions among these waves are given as follows.
\begin{equation}
    \omega_S = \omega_I + \omega
\end{equation}
\begin{equation}
    {\bm k_S} = {\bm k_I} + {\bm k}
\end{equation}
The dispersion relation of the incident and the scattered wave is
\begin{equation}
    \omega^2_{I,S} = k^2_{I,S} c^2 + \omega^2_{pe}.
\end{equation}
The dispersion relations of Langmuir wave, ion acoustic wave,
and beam mode are
\begin{equation}
    \omega^2 = \omega^2_{pe} + k^2 v^2_{te},
\end{equation}
\begin{equation}
    \omega^2 = k^2 C^2_s,
\end{equation}
\begin{equation}
    \omega = k_x u_{bx} + k_y u_{by},
\end{equation}
respectively, where $C_s$ is sound speed.

Substituting eqs.(3)-(5) into eq.(2) gives
\begin{equation}
    k^2_{Sx} c^2 + k^2_{Sy} c^2 \approx (k_I c \pm \omega_{pe})^2.
\end{equation}
Here, $k^2_I c^2 \gg \omega^2_{pe} \gg k^2 v^2_{te}$ has been assumed.
Eq.(8) represents the equations of two circles with radii $k_Ic+\omega_{pe}$
and $k_Ic-\omega_{pe}$, corresponding to the outer and the inner dashed circles
in Fig.\ref{fig:kxky}(d).

Substituting eqs.(3),(4), and (6) into eq.(2), with the assumption
$k^2_I c^2 \gg \omega^2_{pe} \gg k^2 C^2_s$, gives
\begin{equation}
    k^2_{Sx} c^2 + k^2_{Sy} c^2 \approx (k_I c \pm k C_s)^2.
\end{equation}
Note that the above equation does not represent concentric circles
in contrast to eq.(8).
Since $c \gg C_s$, it can be seen that the two circles represented by
the above equation are located approximately midway between the two
dashed circles represented by eq.(8) in Fig.\ref{fig:kxky}(d). The
correspondence between Figs.\ref{fig:kxky}(d) and \ref{fig:kxky}(a) is clear.

Similarly, substituting eqs.(3),(4), and (7) into eq.(2), with the
assumption $k^2_I c^2 \gg \omega^2_{pe} \sim k^2 u^2_b$, gives
\begin{equation}
    k^2_{Sx} c^2 + k^2_{Sy} c^2 \approx (k_I c + k_x u_{bx} + k_y u_{by})^2.
\end{equation}
This leads to an asymmetric circle in Fig.\ref{fig:kxky}(e),
explaining the asymmetric spectrum in Fig.\ref{fig:kxky}(b).
As the beam velocity, $u_b$, decreases and approaches the sound speed,
$C_s$, it nearly overlaps with the inner circle in Fig.\ref{fig:kxky}(d),
making it indistinguishable at this scale (Fig.\ref{fig:kxky}(f)).

%----------
\subsection{CTS spectrum with observed format}
%----------

%
%>>>>>>>>>>>>>>>>>>>>>>>>>>>>>>>>>>>>>>>>>>>>>>>>>>
\begin{figure}[ht]
\includegraphics[clip, width=1.0\columnwidth]{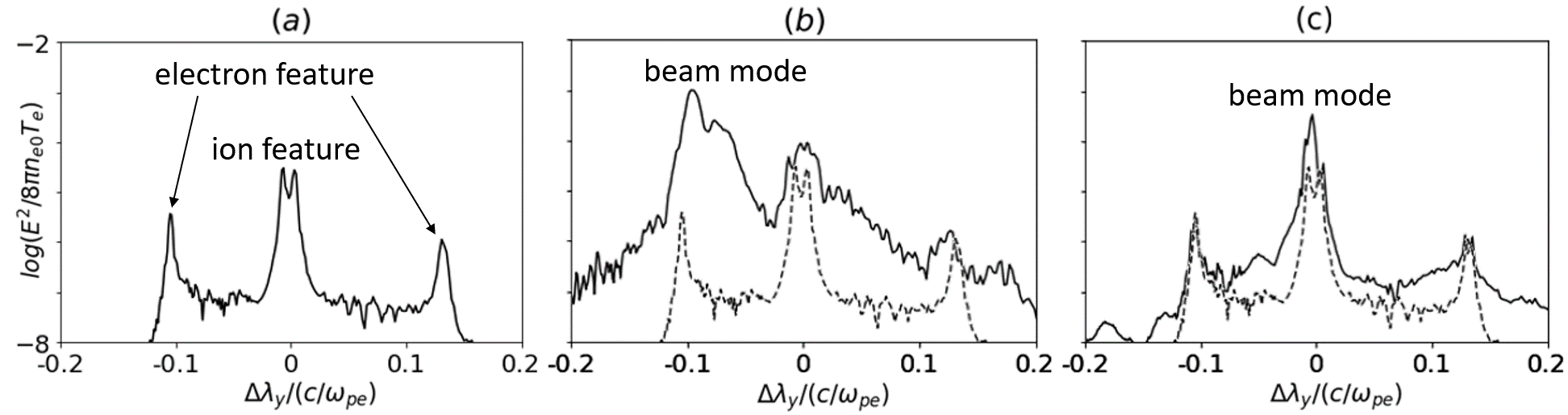}% {fig04.jpg}%Here is how to import EPS art
\caption{\label{fig:I_lambda}Spectra of waves scattered into $-y$-direction
for (a) Run 1, (b) Run 2, and (c) Run 3. The dashed line in (b) and (c)
shows the same data as in (a) for a reference.}
\end{figure}
%>>>>>>>>>>>>>>>>>>>>>>>>>>>>>>>>>>>>>>>>>>>>>>>>>>
%
In a laboratory experiment, the spectrum varies depending on the location of
the scattered wave detection system. As an example, consider the
case where the detection system is placed at -90 degrees relative
to the incident wave (in the negative $k_y$ direction).
Figs.\ref{fig:I_lambda}(a)-(c) show cross-sections of the spectrum
in the $k_y < 0$ region along the $k_y-$axis of
Figs.\ref{fig:kxky}(a)-(c) for Run 1
to Run 3. Here, the horizontal axis represents the $k_y$ values
converted to wavelength and plotted as deviations from the incident
wavelength. The dashed lines in Figs.\ref{fig:I_lambda}(b) and
\ref{fig:I_lambda}(c) are for reference and are the same as those in
Fig.\ref{fig:I_lambda}(a).

In Fig.\ref{fig:I_lambda}(a), characteristic features of the CTS
spectrum of a typical equilibrium plasma can be
observed. The double peak near $\Delta \lambda_y = 0$ corresponds to
the ion feature, while the pair of peaks at both ends represents
the electron feature. The slight asymmetry in the electron feature
peaks arises because the value of $\omega_I / \omega_{pe}$ is smaller
than that in typical experimental conditions due to constraints in
the numerical simulation.

In contrast, in Fig.\ref{fig:I_lambda}(b), the peak corresponding to
the electron feature on the short-wavelength side is significantly
amplified due to the influence of Buneman instability. In
Fig.\ref{fig:I_lambda}(c), the peak corresponding to the ion feature
is enhanced due to ion acoustic instability. Upon closer inspection,
it can also be seen that the ion feature is strongly amplified on
the short-wavelength side. The qualitative spectral features
described above have also been confirmed in 1D simulations
\cite{sakai20,sakai23}.

%
%-------------
\section{\label{sec:weakbeam}CTS in a weak beam-plasma system}
%-------------

\begin{figure}[ht]
\includegraphics[clip, width=1.0\columnwidth]{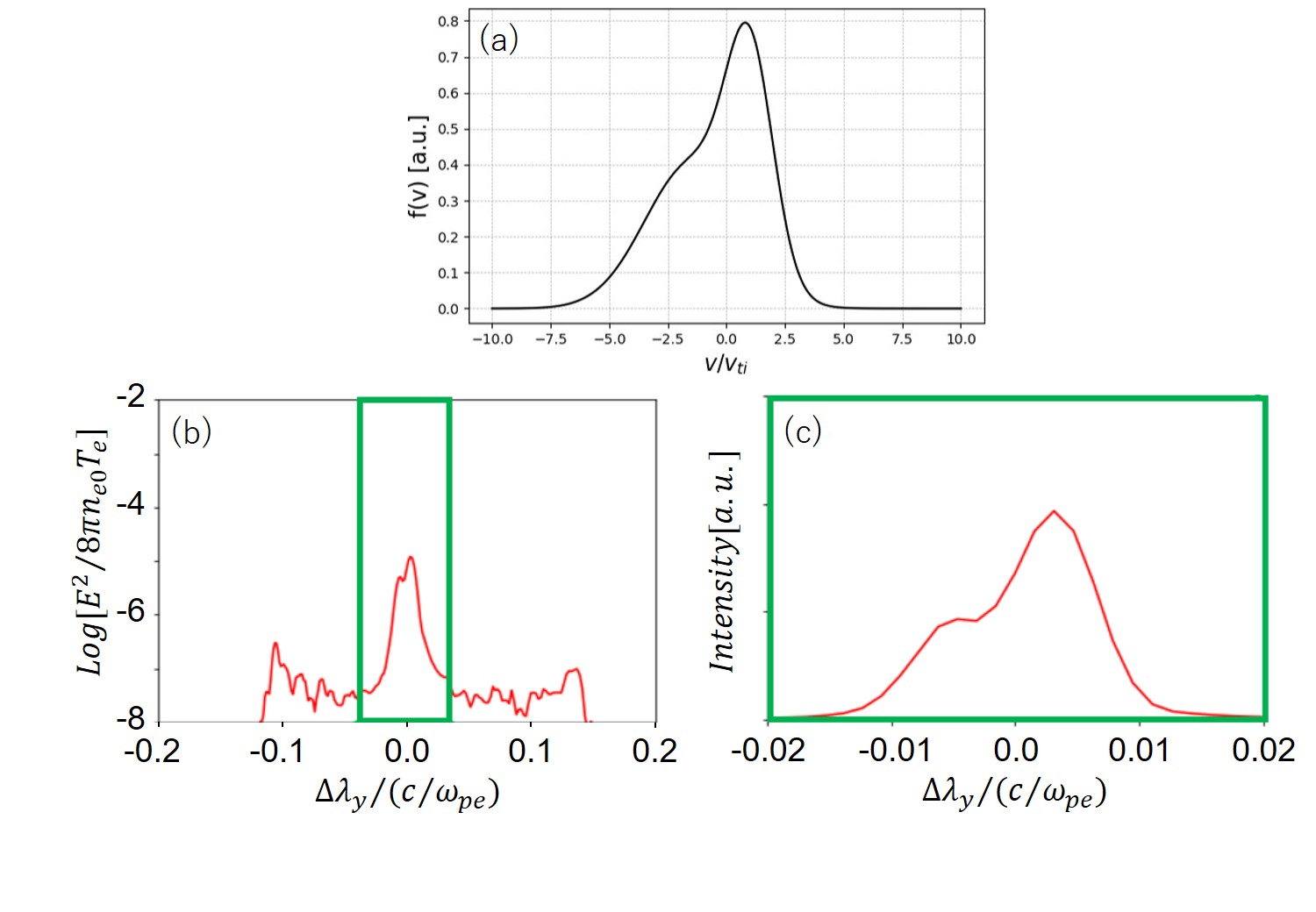}% {fig04.jpg}%Here is how to import EPS art
\caption{\label{fig:weakbeam}(a) Ion distribution function used in
the simulation of the weak beam-plasma system. (b) Simulated scattered
wave spectrum and (c) its expansion (The vertical axis is in a linear scale.).}
\end{figure}

Using this simulation, we can also reproduce CTS spectra for weak beams
that do not lead to instabilities. Here, we consider the case of ion
distribution function shown in Fig.\ref{fig:weakbeam}(a), where the
beam drift velocity is comparable to the thermal
velocity. Since the slope of the distribution function is negative
everywhere, no instability arises. Such distribution functions are often
thought to be realized in quasi-perpendicular shock transition regions
with upstream ion beta of order unity \cite{vvv25}, as well as in
ionospheric plasmas accompanied by ion outflows \cite{wahlund92b}. We set
the beam ion thermal velocity to twice that of the background ion
thermal velocity ($v_{tb}/v_{ti}=2)$, and the electron thermal
velocity to ten times that of the background ions ($v_{te}/v_{ti} =10$),
with using the same mass ratio 25. The relative drift velocity between
the beam ions and electrons is $u_b/v_{ti}=-1.5$, and that between
the background ions and electrons is $u_i/v_{ti}=1$. The ratio of
beam to background ion density is $n_b/n_i=2/3$.

%>>>>>>>>>>>>>>>>>>>>>>>>>>>>>>>>>>>>>>>>>>>>>>>>>>
%>>>>>>>>>>>>>>>>>>>>>>>>>>>>>>>>>>>>>>>>>>>>>>>>>>
%

Fig.\ref{fig:weakbeam}(b) shows the scattered light spectrum obtained in the
same manner as Fig.\ref{fig:I_lambda}. An enlarged view of the ion feature
is shown in Fig.\ref{fig:weakbeam}(c), where the vertical axis is in a
linear scale. An asymmetric ion feature spectrum is reproduced.
%The peak intensity of the ion feature spectrum is similar
%to that in the thermal run (Run 1), while the asymmetric spectral feature
%is reproduced.
A main peak is seen on the longer wavelength side
($\Delta \lambda_y > 0$), with a slightly enhanced signal on the
shorter wavelength side ($\Delta \lambda_y < 0$). The
intensity ratio of this shorter wavelength signal to the main peak
is approximately 0.5.
This asymmetric spectrum in Fig.\ref{fig:weakbeam}(b) and \ref{fig:weakbeam}(c)
can be understood as a scattered wave spectrum produced by two ion
components with different drift velocities and background electrons.

%-------------
\section{\label{sec:discussion}Summary and Discussions}
%-------------

We have self-consistently reproduced collective Thomson scattering in a
nonequilibrium plasma (beam–plasma system) using two-dimensional PIC
simulations and discussed the characteristics of the scattered waves.
In a general two-dimensional in-plane configuration where the beam
velocity and the propagation direction of the incident wave are not
aligned, we examined the wave number spectra of scattered waves
propagating in arbitrary directions. While a symmetric two-dimensional
wave number spectrum is obtained for an equilibrium plasma without a
beam, the beam–plasma system yields an asymmetric spectrum. By taking
into account the linear dispersion relations of the incident wave,
the scattered waves, and the scattering modes (Langmuir waves, ion
acoustic waves, and beam modes), the spectral features can be
largely understood. We further performed simulations for a weak,
linearly stable beam–plasma system with a hot beam, and confirmed
that the obtained scattered wave spectrum shows clear asymmetric
feature.

\textcolor{black}{
The results obtained in this study, as well as those that can be derived using similar numerical approaches, should be useful for interpreting Thomson scattering measurements of nonequilibrium beam–plasma systems that are frequently observed in both space and laboratory plasmas. In the high-latitude ionosphere, beam instabilities are believed to be driven by auroral precipitating electrons, and indeed, Rietveld et al. (1991)\cite{rietveld91} reported asymmetric spectra detected by ISR. In laboratory plasmas as well, recent high-power laser experiments on collisionless shocks often generate localized (ion) beam–plasma systems in the shock transition region and/or developing shock region, where similarly asymmetric scattered wave spectra may arise \cite{vvv25,matsukiyo22,schaeffer19,rinderknecht18,lebedev14}. Although such measurements have traditionally been considered difficult to interpret, we expect that detailed comparison with the numerical data proposed here will enable qualitative and quantitative interpretation. However, several issues still remain, as discussed below.
}

In laboratory experiments and ionospheric radar observations, the frequency of the incident wave is typically about two orders of magnitude higher than the electron plasma frequency. In the present PIC simulations, however, the incident wave frequency is $\omega_I/\omega_{pe}=7.91$, and therefore some care is required for making quantitative arguments. In Sakai et al. (2020\cite{sakai20},2023\cite{sakai23}), the wave equation was solved separately by using electron density fluctuation data obtained from PIC simulations. This approach improves computational efficiency because it avoids solving the incident and scattered waves directly within the PIC simulation. On the other hand, since the nonequilibrium plasma processes and the scattering processes are treated separately, the self-consistency of the model is lost. As seen in Fig.\ref{fig:Ehistory}, in cases where the ratio of the incident wave energy density to the plasma electron thermal energy density is small and plasma heating by the incident wave can be neglected, the method of Sakai et al. (2020\cite{sakai20},2023\cite{sakai23}) should be effective.

%\textcolor{black}{
Related to this, a lower frequency of the incident wave also means a longer incident wavelength. In actual experiments and ionospheric observations, it often happens that $\alpha \equiv 1/k_I \lambda_D$ is on the order of unity, in which case the effects of incoherent scattering cannot be neglected. Milder et al. (2021) \cite{milder21} discuss situations where $\alpha$ crosses unity depending on the scattering angle. In this sense, the present results, corresponding to $\alpha \approx 5.1$, may overestimate the effects of collective scattering.
%}

%In ionospheric observations and laboratory experiments of magnetized plasma shocks, a background magnetic field is present. In section IV, we focused on the internal structure of the shock transition region and therefore assumed that the ions were unmagnetized. On the other hand, electrons are expected to be magnetized, and their effect should in principle be taken into account. However, we believe that their influence on the ion-feature spectrum discussed here is negligible. It is known that when electrons are magnetized, gyro lines appear in the scattered wave spectrum \citep{akbari17}.

In this work, the ion to electron mass ratio was set to 25. Employing the real mass ratio would reduce the sound speed excessively, and reproducing the double-peak structure of the ion feature would then require orders-of-magnitude higher resolution in frequency–wavenumber space, which is not practically achievable. Nonetheless, we consider the present method, capable of self-consistently reproducing CTS in non-equilibrium plasmas, to be highly valuable.

In many space plasma environments and laboratory settings, a background magnetic field is present. However, as long as we focus on electrostatic instabilities driven by field-aligned beams and phenomena occurring on spatial scales smaller than the ion gyro radius, we consider that the influence on the ion feature spectrum discussed here is minimal. %On the other hand, electrons are expected to be magnetized, and examining their effects remains a subject for future work. It is known that when electrons are magnetized, gyro lines appear in the scattered spectrum \citep{akbari17}.
\textcolor{black}{
On the other hand, when electrons are magnetized, the appearance of gyro lines in the scattered wave spectrum has been reported in ionospheric ISR observations.\citep{bhatt06,akbari17} The case presented by Bhatt et al. (2006)\cite{bhatt06} shows a plasma line to gyro line intensity ratio of roughly four. In general, gyro lines are weaker than plasma lines, and long integration times are typically required to achieve a sufficient S/N ratio. However, in situations where gyro lines are efficiently excited via some instabilities, they may be detectable more readily, even with relatively short integration times. Verification of such effects remains a subject for future work.
}

\begin{acknowledgments}
We thank S. Isayama and K. Sakai for fruitful discussions.
This research was partially supported by The Kajima
Foundation for the International Joint Research Grants (2025-06) and JSPS KAKENHI grant No.23K22558 (SM). The computation was carried out using
the computer resource offered under the category of
General Projects by Research Institute for
Information Technology, Kyushu University.
\end{acknowledgments}

\section*{Data Availability Statement}

The data that support the findings of this study are available from the corresponding author upon reasonable request.

\nocite{*}
\bibliography{myaip}% Produces the bibliography via BibTeX.

\end{document}